\documentstyle[preprint,aps]{revtex}
\begin{document}
\title{Phase diagrams of systems with two coupled order parameters}
\author{D. Nicolaides$^{\dagger}$ and A.A. Lisyansky$^{\ddagger}$}
\address{$^{\dagger }$Natural Sciences and Mathematics, Bloomfield College, 467\\
Franklin Street,Bloomfield, NJ 07003\\
$^{\ddagger }$Department of Physics, Queens College of City University of\\
New York, Flushing, NY 11367}
\date{\today}
\maketitle
\tightenlines
\begin{abstract}
Within the framework of an exactly solvable model, which takes into account
the interaction of fluctuating modes with equal and opposite momenta, we
consider phase diagrams in systems with coupled scalar order parameters. We
show that, in agreement with the renormalization group theory, the
fluctuation interaction can split the continuous disorder--order transition
into two phase transitions of the first order. Moreover, the transition may
occur into the anomalous, from the mean field theory point of view phase.
The effect disappears when the fluctuation interaction is suppressed.
\end{abstract}


\section{INTRODUCTION}

Renormalization group (RG) theory made tremendous progress in understanding
the physics of phase transitions of the second order (for review see, e.g., 
\cite{Ma,Aharony,Amit,Baker,ilf-book}). It allows one to calculate critical
exponents with a very high degree of accuracy \cite
{Baker1,Gorishny,Bervillier,Guillon}. Moreover, it predicts a series of
qualitative effects, such as the transformation of a continuous phase
transition in some anisotropic systems into jump-like phase transitions due
to the fluctuation interaction \cite{Aharony,Amit,ilf-book}. Unfortunately,
these transformations cannot be proven rigorously within the framework of
RG. It is just a general belief that a flow of RG trajectories out of the
region of stability of a Ginzburg-Landau-Wilson functional corresponds to a
first order phase transition driven by fluctuations. There is a possibility
that either such a treatment is incorrect or that such a flow away is a
result of approximations, e.g., the $\epsilon $-expansion. Therefore, the
treatment of these systems with the help of exactly solvable models, which
at least partially take into account fluctuation effects, may clarify the
situation. If an exactly solvable model yields results analogous to the ones
of RG theory, it would be a serious argument in support of the reality of
these qualitative effects.

In this paper we consider phase transitions in a system with two coupled
scalar order parameters described by the Ginzburg-Landau-Wilson functional 
\begin{equation}
F(\phi _{1},\phi _{2})=\frac{1}{2}\int d^{3}x\left\{
\sum\limits_{i=1}^{2}\left[ \tau _{i}\phi _{i}^{2}({\bf x})+c_{i}\left(
\nabla \phi _{i}({\bf x)}\right) ^{2}+\frac{1}{4}g_{i}\phi _{i}^{4}({\bf x)}%
\right] +\frac{1}{2}w\phi _{1}^{2}({\bf x})\phi _{2}^{2}({\bf x})\right\} ,
\label{glw}
\end{equation}
where $\tau _{i}=(T-T_{ci})/T_{c1}$, and $T_{ci}$ is a trial critical
temperature for the field $\phi _{i}({\bf x)}$. According to the mean field
theory, lines $\tau _{1}=0$ and $\tau _{2}=0$ on the $(\tau _{1},\tau _{2})$
plane are lines of the second order phase transitions into phases $\phi
_{1}\neq 0,\phi _{2}=0$ and $\phi _{2}\neq 0,\phi _{1}=0$, respectively
[Fig.~\negthinspace 1]\cite{Fisher}. However, the RG analysis shows that
disorder--order phase transitions in the system described by the functional
(1) can be of the first order\cite{Aharony,ilf-book,Kosterlitz,anomalous}.
Our goal is to compare results of the RG theory with results of the exactly
solvable model described below.

Our model uses a simplified physical picture of phase transitions. While
preserving the symmetry of the system, it takes into account interaction of
fluctuations with equal and opposite momenta only. This can be provided by
splitting a $\delta $-function, $\delta ({\bf p}_{1}+{\bf p}_{2}+{\bf p}_{3}+%
{\bf p}_{4})$, which is responsible for the momentum conservation, into a
product of two $\delta $-functions, $\delta ({\bf p}_{1}+{\bf p}_{2})({\bf p}%
_{3}+{\bf p}_{4})$. In coordinate space such a splitting corresponds to the
following reduction of the interaction terms in the functional (\ref{glw}) 
\begin{equation}
\int d^{3}x\phi ^{4}({\bf x)\rightarrow }\frac{1}{V}\left[ \int d^{3}x\phi
^{2}({\bf x)}\right] ^{2},  \label{split}
\end{equation}
where $V$ is the volume of the system. In its simplest version the model has
the same critical asymptotics \cite{Shneider,ilf-tmf} as the spherical model 
\cite{spherical}. However, for more complicated systems it explicitly
demonstrates major qualitative results that have been obtained within the RG
theory, including fluctuation induced first order phase transitions \cite
{ilf-pl,dem-pl} and dimensional crossover for random systems \cite{dem-pr}.

\section{general relations}

In the functional (\ref{glw}) we reduce the interaction terms as follows, 
\begin{eqnarray}
\int d^{3}x\phi _{i}^{4}({\bf x)} &\rightarrow &Va_{i}^{2}[\phi _{i}],
\label{split2} \\
\int d^{3}x\phi _{1}^{2}({\bf x)}\phi _{2}^{2}({\bf x)} &\rightarrow
&Va_{1}[\phi _{1}]a_{2}[\phi _{2}],  \nonumber
\end{eqnarray}
where 
\[
a_{i}[\phi _{i}]=\frac{1}{V}\int d^{3}x\phi _{i}^{2}({\bf x).} 
\]
After reduction (\ref{split2}) the exponent in the partition function
becomes a quadratic form with respect to the functionals $a_{i}[\phi _{i}]$: 
\begin{eqnarray}
Z &\propto &\int D\phi _{1}({\bf x)}D\phi _{2}({\bf x)}  \nonumber \\
&\times &\exp \left\{ -\frac{V}{2}\left[ \sum\limits_{i=1}^{2}\left( \tau
_{i}a_{i}[\phi _{i}]+\frac{1}{V}\int d^{3}xc_{i}\left( \nabla \phi _{i}({\bf %
x)}\right) ^{2}+\frac{1}{4}g_{i}a_{i}^{2}[\phi _{i}]\right) +\frac{1}{2}%
wa_{1}[\phi _{1}]a_{2}[\phi _{2}]\right] \right\} .  \label{partition1}
\end{eqnarray}
Using a transformation analogous to that of Hubbard-Stratanovich, we obtain: 
\begin{eqnarray}
Z &\propto &\int D\phi _{1{\bf q}}D\phi _{2{\bf q}}\int\limits_{-\infty
}^{\infty }dx_{1}dx_{2}\int\limits_{-i\infty }^{i\infty }dy_{1}dy_{2} 
\nonumber \\
&\times &{\bf \exp }\left\{ -\frac{V}{2}\sum\limits_{i=1}^{2}\left[
-iy_{i}\left( x_{i}-a_{i}[\phi _{i}]\right) +\tau _{i}x_{i}+\frac{1}{V}%
\sum\limits_{{\bf q}}c_{i}q^{2}\left| \phi _{i{\bf q}}\right| ^{2}+\frac{1}{4%
}g_{i}x_{i}^{2}\right] +\frac{1}{2}wx_{1}x_{2}\right\} .  \label{partition2}
\end{eqnarray}
We may now perform integrations over all modes except $\phi _{i,{\bf q}=0}$,
which may condense at the point of the phase transition, to obtain 
\begin{eqnarray}
Z &\propto &\int\limits_{-\infty }^{\infty }d\phi _{10}d\phi
_{20}dx_{1}dx_{2}\int\limits_{-\infty }^{\infty }dy_{1}dy_{2}  \nonumber \\
&\times &{\bf \exp }\left\{ -\frac{V}{2}\sum\limits_{i=1}^{2}\left[
x_{i}(\tau _{i}-y_{i})+\frac{1}{4}g_{i}x_{i}^{2}+\frac{1}{4}%
wx_{1}x_{2}+y_{i}\phi _{i0}^{2}+\frac{1}{V}\sum\limits_{{\bf q\neq }0}\ln
|c_{i}q^{2}+y_{i}|\right] \right\} ,  \label{partition3}
\end{eqnarray}
where we define $\phi _{i0}=\phi _{i{\bf q}=0}/\sqrt{V}$. The term $\sum_{%
{\bf q}}\ln |c_{i}q^{2}+y_{i}|$ in Eq.~(\ref{partition3}) diverges on the
upper limit. Strictly speaking, this limit is finite and is equal to the
cutoff momentum of the problem (which, e.g., in solids is a quantity of the
order of the inverse lattice constant). However, critical asymptotics should
not depend upon the cutoff momentum. Therefore, we renormalize sums over $%
{\bf q}$ making them finite and absorb divergencies into renormalizations of
the trial critical temperatures. The renormalization can be done as follows, 
\begin{equation}
\sum_{{\bf q}\neq 0}\!_{R}\ln \left( c_{i}q^{2}+y_{i}\right) =\sum\limits_{%
{\bf q}\neq 0}\ln |c_{i}q^{2}+y_{i}|-\sum\limits_{{\bf q}\neq 0}\ln \left(
c_{i}q^{2}\right) -\sum\limits_{{\bf q}\neq 0}\frac{y_{i}}{c_{i}q^{2}}.
\label{renorm}
\end{equation}
The sum on the left hand side of Eq.~(\ref{renorm}) is convergent even if we
set the upper limit equal to infinity. The second term at the right side is
an unimportant constant, and the third term causes renormalizations of the
trial critical temperatures. As a result we have the partition function in
the form, 
\begin{equation}
Z\propto \int \left( \prod\limits_{i}d\phi _{i0}dx_{i}dy_{i}\right) \exp
\left[ -VF(x_{i},y_{i},\phi _{i0})\right] ,  \label{partition4}
\end{equation}
with function $F$ defined to be, 
\begin{equation}
F(x_{i},y_{i},\phi _{i0})=\frac{1}{2}\sum\limits_{i=1}^{2}\left[
x_{i}(t_{i}-y_{i})+\frac{1}{4}g_{i}x_{i}^{2}+\frac{1}{4}wx_{1}x_{2}+y_{i}%
\phi _{i0}^{2}+\Phi _{i}(y_{i})\right] ,  \label{free}
\end{equation}
where 
\begin{eqnarray*}
\Phi _{i}(y_{i}) &=&\frac{1}{V}\sum_{{\bf q}\neq 0}\/_{R}\ln \left(
c_{i}q^{2}+y_{i}\right) =-\frac{4}{3}\kappa _{i}y_{i}^{3/2}, \\
\kappa _{i} &=&\left( 8\pi c_{i}^{3/2}\right) ^{-1},
\end{eqnarray*}
$t_{i}$ are renormalized critical temperatures, 
\[
t_{i}=\tau _{i}+\frac{g_{i}}{2V}\sum\limits_{{\bf q}\neq 0}\frac{1}{%
c_{i}q^{2}}+\frac{w}{2V}\sum\limits_{{\bf q}\neq 0}\frac{1}{c_{j}q^{2}}%
\equiv \frac{T-T_{i}}{T_{1}}, 
\]
where $i=1,2$ when $j=2,1$.

Due to the fact that $V$ is a large multiplicative constant, one can use the
steepest descent method to calculate the partition function exactly. The
saddle points of the integral (\ref{partition4}) are defined by the
equations, 
\begin{eqnarray}
\frac{\partial F}{\partial x_{i}} &=&\frac{1}{2}\left( t_{i}-y_{i}+\frac{%
g_{i}x_{i}}{2}+\frac{wx_{j\neq i}}{2}\right) =0,  \nonumber \\
\frac{\partial F}{\partial y_{i}} &=&\frac{1}{2}\left( -x_{i}+\phi
_{i0}^{2}-f(y_{i})\right) =0,  \label{saddle} \\
\frac{\partial F}{\partial \phi _{i0}} &=&y_{i}\phi _{i0}=0,  \nonumber
\end{eqnarray}
where 
\[
f(y_{i})\equiv -\frac{d\Phi _{i}}{dy_{i}}=2\kappa _{i}y_{i}^{1/2}. 
\]
Solving this system of equations one can find the equilibrium values of $%
x_{i}$, $y_{i},$ and $\phi _{i0}$. With these values Eq.~(\ref{free})
represents the equilibrium free energy per unit volume of the system.

Eliminating $x_{i}$ from Eqs. (\ref{saddle}) we obtain a simplified system, 
\begin{eqnarray}
t_{i}-y_{i}+\frac{g_{i}}{2}\left[ \phi _{i0}^{2}-f(y_{i})\right] +\frac{w}{2}%
\left[ \phi _{_{j\neq i}0}^{2}-f(y_{_{j\neq i}})\right]  &=&0,
\label{saddle1} \\
y_{i}\phi _{i0} &=&0  \label{saddle2}
\end{eqnarray}
with a free energy defined as 
\begin{eqnarray}
F(\phi _{i0}) &=&\frac{1}{2}\sum\limits_{i=1}^{2}\left\{ t_{i}\left[ \phi
_{i0}^{2}-f(y_{i})\right] +\frac{1}{4}g_{i}\left[ \phi
_{i0}^{2}-f(y_{i})\right] ^{2}+\Phi _{i}(y_{i})\right.   \label{free-e} \\
&&\left. +\frac{1}{4}w\left[ \phi _{10}^{2}-f(y_{1})\right] \left[ \phi
_{20}^{2}-f(y_{2})\right] +y_{i}f(y_{i})\right\} .
\end{eqnarray}

\section{PHASE DIAGRAMS}

According to the mean field theory there are two different types of phase
diagrams for systems with coupled order parameters. They are the diagram
with the bicritical point, where two lines of the second order phase
transitions cross [Fig.~1a], and the tetracritical point diagram that has
four crossed lines of phase transitions of the second order [Fig.~1b]. In
the first case there are two ordered phases, $\phi _{10}\neq 0,\phi _{20}=0$
and $\phi _{10}=0,\phi _{20}\neq 0$, that are separated by a line of the
first order phase transition. At the second case there is an additional
ordered phase, $\phi _{10}\neq 0,\phi _{20}\neq 0$, which is separated from
the other two phases by lines of the second order phase transitions. Below
we reconstruct these diagrams in the framework of the exactly solvable model.

\subsection{ Bicritical point phase diagram}

This diagram is the most interesting one since it can be substantially
modified by the interaction of fluctuations. There are four different
solutions of Eqs.~(\ref{saddle1}) and (\ref{saddle2}). They correspond to
the disordered phase, $\phi _{10}=0,\phi _{20}=0$, and three to the ordered
phases: $\phi _{10}\neq 0,\phi _{20}=0$, $\phi _{10}=0,\phi _{20}\neq 0$ and 
$\phi _{10}\neq 0,\phi _{20}\neq 0.$ In the case of the bicritical point
phase diagram, when $\Delta =g_{1}g_{2}-w^{2}<0,$ only ordered phases
corresponding to the second and the third solutions may occur. Since they
are symmetrical, let us consider, e.g., the phase $\phi _{10}\neq 0,\phi
_{20}=0.$

This solution, according to Eq.~(\ref{saddle2}), requires $y_{1}=0$, $%
y_{2}\neq 0.$ In this case we have: 
\begin{equation}
y_{2\pm }^{1/2}=-\frac{\kappa _{2}\Delta }{2g_{1}}\pm \sqrt{\left( \frac{%
\kappa _{2}\Delta }{2g_{1}}\right) ^{2}+t_{2}-\frac{wt_{1}}{g_{1}}}
\label{y2-1}
\end{equation}
with the following equilibrium values of the order parameter and the free
energy 
\begin{eqnarray}
\phi _{10\pm }^{2} &=&\frac{2}{g_{1}}\left[ -t_{1}-\frac{w\kappa
_{2}^{2}\Delta }{2g_{1}}\pm w\kappa _{2}\sqrt{\left( \frac{\kappa _{2}\Delta 
}{2g_{1}}\right) ^{2}+t_{2}-\frac{wt_{1}}{g_{1}}}\right] ,  \label{order-p-1}
\\
F_{\pm }(\phi _{10}\neq 0,\phi _{2}=0) &=&-\frac{t_{1}^{2}}{g_{1}}+2\kappa
_{2}\left( \frac{wt_{1}}{g_{1}}-t_{2}\right) y_{2\pm }^{1/2}+\frac{\Delta
\kappa _{2}^{2}}{g_{1}}y_{2\pm }+\frac{2}{3}\kappa _{2}y_{2\pm }^{3/2}.
\label{free-e-bi-1}
\end{eqnarray}
When subscripts $1$ and $2$ are interchanged, $1\leftrightarrow 2$, we
obtain solutions for the phase $\phi _{10}=0$, $\phi _{20}\neq 0$. It can be
shown that the free energy $F_{+}$ is always lower than $F_{-}$. Therefore,
only the solution $\phi _{10+}\neq 0$, $\phi _{20}=0$, $\ y_{1+}=0$, $y_{2+}$
$\neq 0$ (or $\phi _{10}=0$, $\phi _{20+}\neq 0$, $\ y_{1+}\neq 0$, $y_{2+}$ 
$=0$) may realize.

According to the mean field theory the phase transition into an ordered
phase is a continues one. However, as one can see from Eq. (\ref{order-p-1}%
), the model predicts that this transition is of the first order if the
following conditions are satisfied 
\[
w>g_{1},\,\,\,\,\,\,w>g_{2},
\]
\begin{equation}
T_{1}\left[ \left( \frac{\kappa _{2}\Delta }{2g_{1}}\right) ^{2}\frac{w-g_{2}%
}{g_{2}}-1\right] >T_{2}\left[ \left( \frac{\kappa _{1}\Delta }{2g_{2}}%
\right) ^{2}\frac{w-g_{1}}{g_{1}}-1\right] ,  \label{1order-condition}
\end{equation}
\[
4g_{1}^{2}(T_{2}-T_{1})+\kappa _{2}^{2}|\Delta
|(w^{2}+g_{1}g_{2}-2wg_{1})T_{1}>0.
\]
They are obtained by requiring that phase $\phi _{10+}\neq 0$, $\phi
_{20}=0\,$has a higher critical temperature than phase $\phi _{10}=0$, $\phi
_{20+}\neq 0$, and that the jump of order parameter $\phi _{10+}$ is
positive.

The transition occurs at the temperature defined by the equation

\begin{equation}
\left( \frac{\kappa _{2}\Delta }{2g_{1}}\right) ^{2}+t_{2}-\frac{wt_{1}}{%
g_{1}}=0  \label{tc1-eq}
\end{equation}
or 
\begin{equation}
T_{c}=\left\{ \left[ \left( \frac{\kappa _{2}\Delta }{2g_{1}}\right) ^{2}+%
\frac{w}{g_{1}}\right] T_{1}-T_{2}\right\} \left( \frac{w}{g_{1}}-1\right)
^{-1}.  \label{tc1}
\end{equation}
The jump of the order parameter at the point of transition is equal 
\[
\phi _{10}^{2}(T_{c})=\frac{2}{g_{1}}\frac{T_{1}\left[ 1+\left( \frac{\kappa
_{2}\Delta }{2g_{1}}\right) ^{2}\right] -T_{2}}{T_{1}\left( \frac{w}{g_{1}}%
-1\right) }+\frac{\kappa _{2}w\left| \Delta \right| }{g_{1}^{2}}. 
\]
Such a transition occurs not necessarily into the phase with a higher trial
critical temperature. Comparing free energies for the phases $\phi _{10}\neq
0$, $\phi _{20}=0$ and $\phi _{10}=0,\phi _{20}\neq 0$ one can show that for 
\begin{equation}
T<T_{c}\text{ \qquad and \qquad }T_{1}<T_{2},  \label{anomal}
\end{equation}
and when conditions of inequalities (\ref{1order-condition}) are true,
\thinspace the lower energy corresponds to the ``anomalous'', from the mean
field theory point of view, phase $\phi _{10}\neq 0$, $\phi _{20}=0$. Only
if $T_{2}<T_{1}$ the transition happens into the ``normal'' phase, $\phi
_{10}\neq 0$, $\phi _{20}=0$.

The analogous result has been obtained in the RG\ theory \cite
{ilf-book,anomalous} where it was shown that in the system with competing
order parameters, when $\Delta <0$ and trial critical temperatures, $T_{1}$
and $T_{2}$, are close to each other, the transition into the ordered phase
is of the first order into the anomalous phase with the lower trial critical
temperature. Such a phase is stabilized by the interaction of critical
fluctuations, and when the temperature decreases further another phase
transition into the ``normal'' phase must occur. Since the transition into
the ``normal'' phase is the transition of the order-order type it must also
be of the first kind. Hence, the fluctuation interaction splits the
continuous phase transition, predicted by the mean field theory, into two
transitions of the first order. The effect must disappear when fluctuations
are suppressed. The model provides a simple way of regulating the strength
of fluctuations. Fluctuations are suppressed in systems with long-range
interactions. The parameters $c_{i}^{1/2}$ represent the radii of
interactions in the original system in which critical behavior is described
by Ginzburg-Landau functional (\ref{glw}). The Ginzburg number, $Gi,$ that
defines the width of the domain where fluctuation are essential, is
proportional to $c_{i}^{-3}$ or $\kappa _{i}^{2}$. So, setting $\kappa
_{i}\rightarrow 0$ we suppress fluctuations. In this case the order
parameter (\ref{order-p-1}) becomes proportional to $\sqrt{-t_{1}}$ and the
transition becomes of the second order. It is interesting to note that the
effect of inducing the first order transition into the phase $\phi _{10}\neq
0$, $\phi _{20}=0$ is controlled only by the fluctuations of the order
parameter $\phi _{20}$ (and, of course by the coupling, $w$, between modes).
Using the model, the phase diagram corresponding to the first order phase
transition into the anomalous phase is schematically shown in Fig.~2.

\subsection{Tetracritical point phase diagram}

In the mean field theory the case of $\Delta >0$ corresponds to the phase
diagram with the tetracritical point shown in Fig.~1b. According to the
model, the phase $\phi _{10}\neq 0$, $\phi _{20}=0$ corresponds to the
solution of Eqs.~(\ref{saddle1}) and (\ref{saddle2}). The result is $\phi
_{10+}$ defined by Eq. (\ref{order-p-1})\thinspace , which independently of
the sign of $w$, always has a solution and the transition is of the second
order. The critical temperature in this case is defined by 
\begin{equation}
T_{c}=T_{1}+\frac{w\kappa _{2}^{2}(w-g_{2})T_{1}}{2}+\frac{w\kappa _{2}T_{1}%
}{2}\sqrt{\kappa _{2}^{2}(w-g_{2})^{2}+\frac{4(T_{1}-T_{2})}{T_{1}}.}
\label{Tc-2}
\end{equation}

The RG\ theory, however, allows for a fluctuation induced first order phase
transition when $\Delta >0$ and $w<0$. This discrepancy can be attributed to
that the model takes into account fluctuations in a limited manner.

The transition between ordered phases is described by Eqs.~(\ref{saddle1})
and (\ref{saddle2}) with $y_{1}=$ $y_{2}=0:$%
\begin{eqnarray}
t_{1}+\frac{g_{1}}{2}\phi _{1}^{2}+\frac{w}{2}\phi _{2}^{2} &=&0,  \nonumber
\\
t_{2}+\frac{g_{2}}{2}\phi _{2}^{2}+\frac{w}{2}\phi _{1}^{2} &=&0.
\label{mixed-eq}
\end{eqnarray}
These equations have ordinary mean field theory solutions 
\begin{eqnarray}
\phi _{10}^{2} &=&2(wt_{2}-g_{2}t_{1})/\Delta ,  \nonumber \\
\qquad \phi _{20}^{2} &=&2(wt_{1}-g_{1}t_{2})/\Delta ,  \label{mixed}
\end{eqnarray}
which exists when 
\[
\Delta >0,\qquad wt_{2}-g_{2}t_{1}>0,\qquad wt_{1}-g_{1}t_{2}>0.
\]
According to Eqs.~(\ref{mixed}) in agreement with the mean field theory
transitions between ordered phases for positive $\Delta $ are of the second
order. Besides the renormalization of critical temperatures, according to
the model, fluctuations do not change the tetracritical point phase diagram.

\newpage

\section*{Figure Captions}

Fig. 1. Phase diagrams with bicritical (a) and tetracritical (b) points in
the mean field theory. For these and the next figure dashed lines correspond
to continuous phase transitions, the solid line corresponds to a phase
transition of the first order between two ordered phases.

Fig. 2. Bicritical point phase diagram modified due to fluctuations. Near
the point where lines of the order-disorder transitions cross all phase
transitions become of the first order. With decreasing temperature the state
of the system moves along the dotted line. When this line crosses the first
solid line, as shown, the disorder-order transition is of the first order
into the ``anomalous'' phase instead of the continuous phase transition
described by the mean field theory. When the dotted line crosses the second
solid line the first order transition into the ``normal'' phase occurs.

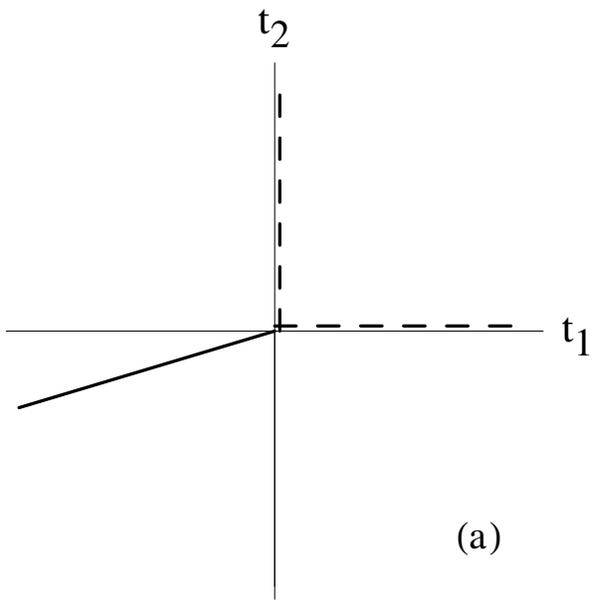
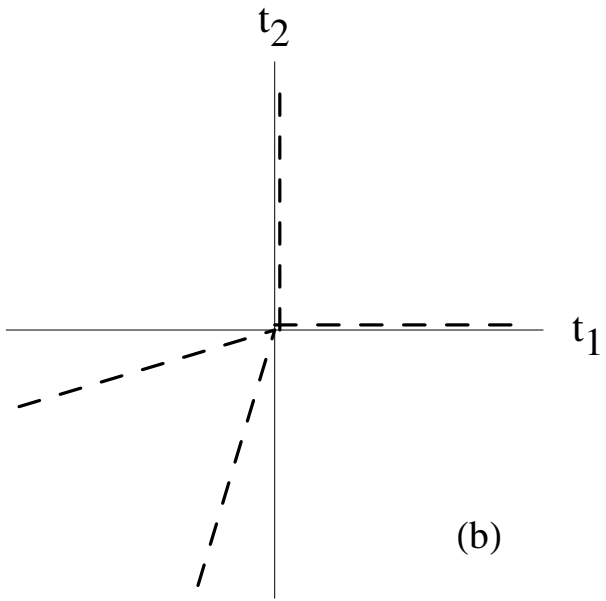
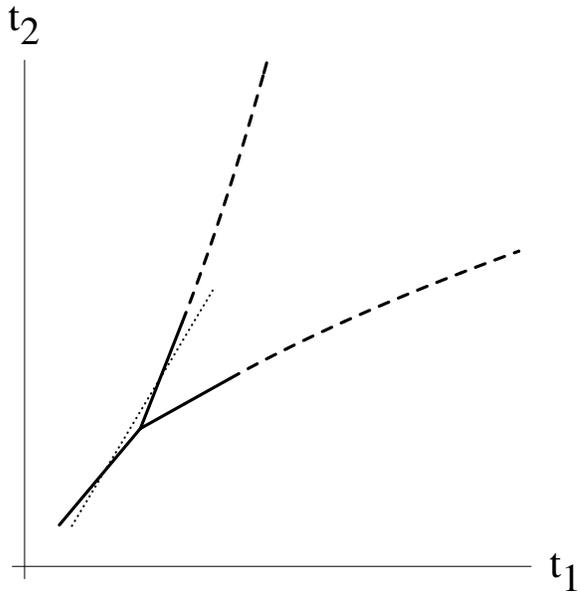
\end{document}